# Active Dual-Gated Graphene Transistors for Low-Noise, Drift-Stable, and Tunable Chemical Sensing

Vinay Kammarchedu[1,2,3], Heshmat Asgharian[1,3], Hossein Chenani[1,3], and Aida Ebrahimi[1,2,3,4,5,*]

[1]Department of Electrical Engineering, [2]Center for Atomically Thin Multifunctional Coatings,
[3]Materials Research Institute, [4]Department of Biomedical Engineering,
[5]Center for Neural Engineering,
The Pennsylvania State University, University Park, Pennsylvania 16802, United States

*Corresponding author: sue66@psu.edu

## Abstract

Graphene field-effect transistors (GFETs) are among the most promising platforms for ultrasensitive chemical and biological sensing due to their high carrier mobility, large surface area, and low intrinsic noise. However, conventional single-gate GFET sensors in liquid environments suffer from severe limitations, including signal drift, charge trapping, and insufficient signal amplification. Here, we introduce a dual-gate GFET architecture that integrates a high-κ hafnium dioxide local back gate with an electrolyte top gate, coupled with real-time feedback biasing. This design enables capacitive signal amplification while simultaneously suppressing gate leakage and low-frequency noise. By systematically evaluating seven distinct operational modes, we identify the Dual Mode Fixed configuration as optimal, achieving up to 20× signal gain, > 15× lower drift compared with gate-swept methods, and up to 7× higher signal to noise ratio across a diverse range of analytes, including neurotransmitters, volatile organic compounds, environmental contaminants, and proteins. We further demonstrate robust, multiplexed detection using a PCB-integrated GFET sensor array, underscoring the scalability and practicality of the platform for portable, high-throughput sensing in complex environments. Together, these advances establish a versatile and stable sensing technology capable of real-time, label-free detection of molecular targets under ambient and physiological conditions, with broad applicability in health monitoring, food safety, agriculture, and environmental screening.

## 1. Introduction

Graphene field-effect transistors (GFETs) have emerged as exceptionally sensitive platforms for chemical and biological sensing [1–5]. The atomically thin graphene channel offers a fully exposed surface and exceptional carrier mobility, enabling strong field-effect responses to adsorbates [6]. In typical single-gate GFET biosensors, baseline stability and limit-of-detection (LOD) are degraded by several well-known issues [7–9]. One major problem is signal drift and hysteresis: the graphene transfer characteristics (e.g., Dirac point) tend to shift irreversibly or slowly over time (in some cases even exacerbated by applied electrostatic gates) due to charge trapping and adsorbates [10–12]. Likewise, substantial hysteresis is seen in graphene-on-oxide devices: as the gate voltage is swept, charge transfer and capacitive effects (from adsorbed molecules or ions) cause positive or negative voltage shifts of the conductance minimum [13–16]. Unexplained signal drift and interfacial phenomena at the nanoscale remain key obstacles to the stable operation of graphene-based FET biosensors [7].

These performance-limiting instabilities are further compounded by the reliance on single-gate architectures (majority of reported GFET research) – typically employing an electrolyte top gate for liquid-phase analyte detection or a back gate for solid-state or gaseous phase sensing [17]. This static-gate mode of operation often yields relatively small signal amplitudes, limiting sensitivity and signal-to-noise ratio. To enhance the sensing response, many studies employ dynamic gate sweeps to probe features such as the position of the Dirac point, transconductance, and carrier mobility – parameters that are modulated by analyte interactions [17, 18]. Gate voltage sweeps can induce charge trapping/detrapping, hysteresis, and temporal signal drift. To overcome the limitations associated with small signal amplitudes and drift-prone gate sweeping, there is a pressing need for signal amplification strategies that do not compromise measurement stability or noise performance.

Dual-gate transistor architectures present a compelling alternative by enabling capacitive signal amplification through asymmetric gate coupling [19–22]. In dual-gate transistors, the channel is modulated simultaneously by two independent gate electrodes – for example, a solid-state dielectric back gate and an electrolyte top gate. When these gates are designed with significantly different gate capacitances – typically with the electrolyte gate providing orders of magnitude higher capacitance than the back gate – a small perturbation at the top gate (e.g., due to analyte binding) can result in a disproportionately larger signal response when the back gate compensates via capacitive coupling [22]. This asymmetric dual-gating framework enables signal amplification without the need for rapid gate sweeping. This concept has been previously explored in a limited number in non-graphene sensing platforms. For example, a dual-gated silicon nanowire FET, incorporating both a liquid gate and a back gate, demonstrated pH sensitivity beyond the Nernst limit [21]. More recently, another study utilized a dual-gated GFET with an ionic liquid top gate and an actively controlled back gate under high-vacuum conditions to enhance pH sensitivity via feedback mechanisms [22]. In this configuration, the graphene channel was isolated from the analyte medium and served solely as an electronic amplifier within the feedback loop.

Despite these promising reports, dual-gated GFET sensors remain underexplored, particularly those that fully leverage graphene's intrinsic surface sensitivity and chemical tunability – whether through direct functionalization with molecular probes or through electrolyte-gated interactions with charged analytes in solution. A key challenge for dual-gate GFETs is gate leakage in liquid environments. While the use of local high-κ back gates has been shown to substantially reduce effective oxide thickness (EOT) and suppress leakage, these benefits have largely been demonstrated under dry conditions [23–25]. However, in a liquid environment, the back-gate electrode area must be carefully minimized to avoid faradaic currents due to defects in the oxide [26–28]. We believe this challenge is a key reason why dual-gate GFETs have not been widely adopted in research or industry.

In this work, we overcome longstanding limitations of GFET sensors by introducing a field-effect amplified, active dual-gated architecture engineered for enhanced sensitivity and operational stability with low noise. Our design integrates a local high-κ hafnium dioxide ($HfO_2$) back gate, patterned beneath the graphene channel, paired with a liquid-phase electrolyte top gate. This asymmetric dual-gate configuration acts as a capacitive voltage divider: perturbations at the graphene-electrolyte interface – such as ion adsorption or analyte binding – modulate the surface potential, which is then amplified at the back gate via capacitive coupling. To achieve active feedback, we employ readily available operational amplifiers to dynamically control the back gate voltage in real time. This approach enables cost-effective, on-demand modulation of the gate potential, facilitating sensor calibration and drift compensation without requiring complex electronics.

To systematically evaluate sensing performance, we define and compare seven distinct operational modes, encompassing a range of static and dynamic biasing conditions, both in single- and dual-gate configurations. Each mode was assessed across various analyte categories, including pH changes, small molecules (electroactive biogenic neurotransmitter amines), volatile organic compounds (isopropyl alcohol), environmental contaminants (PFAS), and large biomolecules (cytokine IL6). This wide applicability underscores the versatility of the sensing platform. Among all configurations, the Dual Mode Fixed (DMF; described in later sections) mode demonstrates the highest signal fidelity and sensor performance. In terms of performance metrics, DMF achieves up to 20× times higher sensitivity, > 15× lower signal drift, and enhanced signal-to-noise ratios (up to 7×) – well beyond what is attainable using conventional single-gate GFETs.

In summary, the proposed dual-gated, field-effect amplified GFET architecture establishes a new standard for graphene-based sensors by simultaneously delivering ultrasensitive detection, ultralow drift, and on-demand signal amplification. By integrating advanced materials and device engineering with tailored electronic circuit design and rigorous operational mode analysis, this platform achieves unprecedented sensor performance. Moreover, the approach is broadly transferable to other 2D materials and nanoscale FET systems, providing a versatile blueprint for next-generation biosensors and environmental monitors requiring both exceptional sensitivity and long-term stability. We envision this platform enabling real-time, high-precision sensing in complex environments, with the potential to achieve single-molecule detection limits in portable, low-power devices.

## 2. Device Architecture and Signal Amplification Principle

The device architecture utilized in this study is a generalized dual-gated GFET, incorporating two independent gating mechanisms: (1) a top gate formed by an electrolyte interface and (2) a local solid-state back gate. A schematic representation of the structure is shown in Fig. 1a (Inset shows a representative fabricated device; as a comparison, similar schematic for global gated devices shown in Fig. 1b). This dual-gate configuration enables versatile modulation of the graphene channel potential and charge carrier density via capacitive coupling from both gates [29].

To model the operation of the dual-gate GFET, we employed a standard electrostatic model involving two capacitances in series: the geometric capacitance of the gate dielectric and the quantum capacitance of graphene [30]. The quantum capacitance arises due to the linear dispersion relation of Dirac fermions in graphene, which leads to a low density of states near the Dirac point. The total gate capacitance for each gate, $C_{eff}^{(i)}$, is thus given by Equation 1:

$$\frac{1}{C_{eff}^{(i)}} = \frac{1}{C_{geo}^{(i)}} + \frac{1}{C_q} \quad \text{Eq. 1}$$

, where $i = \{TG, BG\}$ denotes top and back gate, $C_q$ is the quantum capacitance of graphene, and $C_{geo}^{(i)}$ is the geometric capacitance for the top gate and back gate. For the top gate, the geometric capacitance is the double layer capacitance, $C_{dl}$, formed by the electrolyte and is heavily influenced by the ionic strength, media type, and dissolved species. For the back gate, the geometric capacitance is the oxide capacitance, $C_{ox}$. Using this model, the drain source current of this field effect transistor is modeled by Equation 2:

$$I_{DS} = \mu \frac{W}{L} V_{DS} \left( C_{eff}^{TG}(V_{TG} - V_{TG,Dirac}) + C_{eff}^{BG}(V_{BG} - V_{BG,Dirac}) + en' \right) \quad \text{Eq. 2}$$

, where $\mu$ is the carrier mobility, $W$ and $L$ are width and length of the graphene channel, $e$ is the electron charge, $n' = n_0'$ is intrinsic doping including trapped charges and charge puddles, $V_{DS}$ is the drain source voltage, $V_{TG}$ and $V_{BG}$ are the top and back gates respectively, $V_{TG,Dirac}$ and $V_{BG,Dirac}$ are the Dirac peak locations for top and back gate respectively. Interaction of graphene and analyte as a measurement progress can be modeled as changes in $n' = n_0' + \Delta n' + a$ in Equation 2. Hence, in a GFET at a fixed gate potential,

$$\Delta I_{DS} = \mu \frac{W}{L} eV_{DS}(\Delta n' + a) \quad \text{Eq. 3}$$

, where $\Delta n'$ is the added drift/hysteresis due to the performed measurement, $\Delta I_{DS}$ is the signal change, and $a$ is the contribution from charge introduced due to the analyte-graphene/gate interaction. If we assume the noise in $I_{DS}$ to be $N_{I,DS}$, the signal to noise ratio (SNR) – when $\Delta I_{DS}$ is used as the signal – using Equation 3 can be calculated as,

$$SNR_{I,DS} = \frac{a}{\Delta n' + \frac{N_{I,DS}}{\mu \frac{W}{L} eV_{DS}}} \quad \text{Eq. 4}$$

If the gate potential is fixed for entirety of the measurement, we can assume $a$ is much larger compared to $\Delta n'$, as is the case where measurements commonly bias the gate and continuously read $I_{DS}$. Thus, $SNR_{I,DS}$ of change in these cases is mostly proportional to $a/N_{I,DS}$. In contrast, if $a$ is small/comparable to $\Delta n'$, $SNR_{I,DS}$ is small, as is the case with measurements where the gate is swept rapidly. In these measurements other metrics such as the location of the Dirac peak is used instead, ideally this is equal to:

$$\Delta V_{DP,(i)} = \frac{e(\Delta n' + a)}{C_{eff}^{(i)}} \quad \text{Eq. 5}$$

, where $\Delta V_{DP,(i)}$ is the change in the Dirac peak location due to change in $\Delta n'$. Similar to Equation 4, we can calculate the SNR for Equation 5 – assuming $N_{V,DP,(i)}$ as the noise in $V_{DP,(i)}$ – as:

$$SNR_{V,DP,(i)} = \frac{a}{\Delta n' + \frac{C_{tot}^{(i)}}{e} N_{V,DP,(i)}} \quad \text{Eq. 6}$$

Generally, in order to find the Dirac peak, multiple $I_{DS}$ measurements are performed (assume $N$). This means that we can estimate (from Equation 2) that,

$$C_{eff}^{(i)} N_{V,DP,(i)} \times f(N) = \frac{N_{I,DS}}{\mu \frac{W}{L} e V_{DS}}$$  Eq. 7

, where $f(N) \propto \sqrt{N}$ is a function of $N$ guaranteed to be greater than unity since multiple measurements reduce the uncertainty [31]. Hence, from Equations 4, 6 and 7, $SNR_{V,DP,(i)} > SNR_{I,DS}$ provided $\Delta n'$ does not increase substantially enough to overshadow the decrease in measurement uncertainty due to the sweeps as compared to the fixed gate method, this is generally the case for slow sweeps.

For a simultaneous dual gating with feedback that compensates for changes in $I_{DS}$ by sweeping the back gate, we require that ideally $\Delta I_{DS} = 0$. Hence, using Equation 2, we have,

$$C_{eff}^{TG} \Delta V_{TG} + \Delta V_{BG} C_{eff}^{BG} + e\Delta n' + ea = 0$$  Eq. 8

Which follows that the measured signal $\Delta V_{BG}$ is given by,

$$\Delta V_{BG} = -\frac{(e\Delta n' + ea + C_{eff}^{TG} \Delta V_{TG})}{C_{eff}^{BG}} = -\frac{C_{eff}^{TG}}{C_{eff}^{BG}} \Delta V_{e,TG}$$  Eq. 9

, where $\Delta V_{e,TG} = e \frac{\Delta n' + a}{C_{eff}^{TG}} + \Delta V_{TG}$ is the equivalent top gate shift due to either molecule interaction with gate or graphene channel. Hence, following this analysis, for simultaneous dual gating – where we design a feedback system to fix $I_{DS}$ by actively changing $V_{BG}$ and fixing $V_{TG}$ – we find that Equation 5 still holds, with the signal defined as:

$$\Delta V_{BG} = \frac{e(\Delta n' + a)}{C_{eff}^{BG}}$$  Eq. 10

Similarly, from Equation 7, we preserve the gains in SNR. Moreover, since the sweep rate of $V_{BG}$ is practically zero, we can assume $\Delta n'$ is also much lower.

In conclusion, we find that compared to fixed gate systems, the signal with our dual-gating approach is amplified by $C_{eff}^{TG}/C_{eff}^{BG}$ (from Equation 9), while SNR is also improved since lower speed gate sweep rates lead to less $\Delta n'$. This necessitates that we estimate both $C_{eff}^{TG}$ and $C_{eff}^{BG}$ to calculate the expected gain in sensitivity. From literature review (and experimental evidence in Section 3.3) we find that $C_{geo}^{TG} = C_{dl} > C_q$ and $C_{geo}^{BG} = C_{ox} \ll C_q$. Accordingly, Equation 9 can be approximated to:

$$\frac{\Delta V_{BG}}{\Delta V_{e,TG}} = -\frac{C_{dl} C_q}{C_{ox}(C_{dl} + C_q)}$$  Eq. 11

Using approximate values for $C_{dl} \approx 36 \frac{\mu F}{cm^2}$, $C_q \approx 7 \frac{\mu F}{cm^2}$, and $C_{ox} \approx 0.6 \frac{\mu F}{cm^2}$ in our system, we estimated a gain of approximately $\frac{\Delta V_{BG}}{\Delta V_{e,TG}} \approx 10$ (see Section S3). However, as these capacitances change – especially $C_{dl}$ due to changes in media or $C_q$ due to changes in the media, molecule interaction, or gate potentials – we expect to see varied amplification factors. Since $C_q$ is slightly smaller than $C_{dl}$, it plays a significant role in determining the amplification factor as described in Section 3.3.

## 3. Operational Modes and Performance Comparison

The device fabrication and experimental validation were conducted using various electrolyte media and analytes, as described in Section 5.1. The local back gate consisting of a thin $HfO_2$ dielectric is deposited over patterned electrodes on a silicon dioxide substrate, as described in Section 5.2 and schematically shown in Fig. S1 of Section S1. Briefly, photolithography and atomic layer deposition (ALD) were performed on a $Si/SiO_2$ wafer to pattern a tri-metal local gate, deposit 35 nm $HfO_2$ dielectric layer, and etch openings for the source and drain contacts. Graphene was then transferred onto the local gate stack via a wet transfer method, followed by etching and passivation to complete device fabrication, as detailed in Section 5.3 and illustrated in Fig. S2. The finalized device stack was integrated with the measurement readout circuitry, described in Section 5.4 (shown in Section S2) and tested using protocols for various media and analytes outlined in Section 5.5.

The dual-gated GFET architecture facilitates the exploration of multiple operational modes by independently configuring the top and back gate voltages. These modes include single-gate operation (with one gate floated or fixed), dual-sweep operation (both gates swept), and differential feedback operation. Table 1 summarizes the seven operational modes evaluated in this study.

### 3.1 Single-Gate Operation Modes: Top Gate Fixed (TGF) and Top Gate Sweep (TGS)

In Top Gate Fixed (TGF) mode, the top gate is held at a constant voltage while the back gate is left floating, as illustrated in Fig. 1c. In this configuration, the device functions effectively as a single top-gated GFET, where the electrolyte forms an electric double layer (EDL) at the graphene-electrolyte interface. This EDL serves as a high-capacitance gate dielectric, enabling modulation of the graphene channel at ultra-low gate voltages. The elevated EDL capacitance enhances charge carrier accumulation in the channel, facilitating real-time monitoring of the drain current during analyte exposure. In the Top Gate Sweep (TGS) mode, the top gate voltage is swept while the back gate remains floating, as illustrated in Fig. 1d. This approach enables acquisition of full transfer characteristics, allowing extraction of transconductance and precise tracking of Dirac point shifts caused by surface interactions. These shifts result from electrostatic gating by target analytes or specific molecular interactions at the graphene surface or gate electrode. For example, adsorption of charged biomolecules such as DNA or proteins onto the graphene channel induces local doping, causing concentration-dependent shifts of the Dirac point. Furthermore, reactions or binding events at the gate-electrolyte interface – such as protonation of functional groups – can indirectly modulate the gating field experienced by the graphene channel [9].

The nature and strength of the electrolyte play a significant role in device behavior. Electrolytes with smaller ions or higher ionic strength can form thinner and more compact EDLs, resulting in higher effective gate capacitance [32]. Aqueous electrolytes – such as phosphate-buffered saline and ionic liquids – enable strong gating but can differ markedly in their dielectric behavior, viscosity, and ion mobility. These differences directly affect the position and sharpness of the

Dirac point. Specifically, weaker electrolytes or those containing bulky solvated ions – such as organic solvents like acetonitrile – result in broader or shifted transfer characteristics due to the formation of a more diffuse EDL. Table S1 summarizes the extracted conductance and double layer capacitance for different media, including deionized water (DIW), phosphate-buffered saline (PBS), potassium chloride (KCl), acetonitrile (ACET) with and without 10 mM KCl, and an ionic liquid (IL).

As expected, ionic media such as PBS and 10 mM KCl exhibited high capacitance (36-46 $\mu F/cm^2$) and low solution resistance (~0.2 k$\Omega$), consistent with efficient ionic screening. In contrast, DIW and pure acetonitrile showed low capacitance in the sub-pF/cm² range with much higher solution resistance, reflecting their low ionic conductivity. The ionic liquid demonstrated intermediate capacitance and resistance, highlighting its unique electrochemical properties. These results validate the strong dependence of interfacial capacitance on electrolyte composition, which is critical for sensor performance in different media. Section S3 calculates the double layer capacitance (for PBS) of our device to be around 1.4 nF and the back gate capacitance to be around 38 pF, providing an approximate ratio of ~36. As described in Section 2, this ratio dictates the potential gain from the feedback system. However, due to inaccuracies with double layer capacitance measurement, changing quantum capacitance of graphene, along with other geometric considerations we find that in our system this ratio is ~10 (experimentally verified in Section 3.4).

Example TGS data of devices measured in PBS is shown in Fig. 1e displaying the characteristic Dirac peak of graphene which shifts as multiple gate scans are performed. A distribution of the initial resistance at zero gate ($R_{SD,0}$), ratio of resistance at the Dirac peak to $R_{SD,0}$ ($R_{Mod}$), and the location of the Dirac peak ($V_{DP}$) for 18 devices also individually plotted in Fig. S4.

### 3.2 Back-Gate Operation Modes: Back Gate Fixed (BGF) and Back Gate Sweep (BGS)

In Back Gate Fixed (BGF) mode, the back gate voltage is held constant while the top gate is floated, allowing partial investigation of solid-state gating contributions, as shown in Fig. 1f. In contrast, Back Gate Sweep (BGS) mode involves actively sweeping the back gate while the top gate remains floated, as shown in Fig. 1g. These configurations help isolate and characterize the intrinsic behavior of the device structure, particularly the effects of solid-state dielectric modulation in the absence of an electrolyte interface.

A key challenge with back gate operation – especially in ambient conditions – is the inherent high doping of graphene due to adsorbed moisture, oxygen, and charged species from the environment [15, 16]. This ambient doping often masks or shifts the Dirac point, making precise calibration difficult. Moreover, solid-state gating in thin dielectric systems is prone to leakage currents and potential dielectric breakdown, especially under high electric fields. These effects can compromise long-term device reliability.

However, in our devices fabricated using ALD of $HfO_2$ as the back gate dielectric, we observe clear and reproducible Dirac peaks at significantly lower back gate voltages, as shown in Fig. 1h. This contrasts sharply with devices using, for example, 285 nm $SiO_2$ oxide which require back gate voltages exceeding 80V [15, 16]. This improvement is attributed to the higher dielectric constant (k ~25) of $HfO_2$, resulting in approximately a 50× increase in capacitance, which enables efficient capacitive coupling at reduced gate biases. Compared to conventional global back-gated $SiO_2$ devices (Fig. 1b), our architecture delivers superior performance. The lower geometric capacitance of thick $SiO_2$ (~50× smaller than that if our devices) limits its suitability for low-voltage or battery-powered applications. Additionally, careful processing and cleaning minimize trapped charges and charge puddles, further stabilizing device operation. Overall, across tested sensors, we achieved a 94% yield for successful source-drain contacts and graphene presence under dry conditions. However, only 65% of devices exhibited functional back-gate response. Electrical gate leakage failures accounted for 35% of devices, while resistive failures were observed in 6%. Distribution of device parameters are plotted as insets in Fig. 1h and detailed in Section 3.1. Individual device data for 63 devices are presented in Fig. S5.

From a system integration perspective, global back-gated configurations – such as those employing a common $SiO_2$ back plane – are less suitable for array-based sensing. Due to the fabrication variability inherent in two-dimensional materials, a single leaky device in a globally gated array can affect all sensors sharing the same gate, compromising measurement integrity. In contrast, our locally patterned back gate architecture provides device-level control, enabling selective gating and isolation of individual sensors. This localized gating approach enhances robustness, improves fault tolerance, and offers clear advantages for multiplexed biosensing and scalable integration.

### 3.3 Addition Mode (AM)

In the Addition Mode (AM), both the top electrolyte gate and the back solid-state gate are sequentially swept during a single measurement, as illustrated in Fig. 2a. This dual-gate operation enables complex, high-resolution mapping of the GFET transfer characteristics, providing detailed insights into charge neutrality and electrostatic coupling across the graphene channel. Notably, to the best of our knowledge, this is the first demonstration of a dual-gated GFET system in which both a solid oxide back gate and a liquid/aqueous electrolyte top gate yield clearly resolved Dirac peaks under simultaneous sweep conditions.

Previous studies have typically relied on asymmetric gate structures in which one gate – often the electrolyte – dominates electrostatic control, rendering the secondary gate largely ineffective in modulating the channel. In contrast, our devices achieve comparable gate strengths between the top and back gates, enabled using high-κ $HfO_2$ as the back-gate dielectric. This balance allows both gates to independently and effectively modulate carrier concentration, resulting in distinctly measurable Dirac peaks from each gate – an unprecedented result for GFETs.

We must note that repeated back gate sweeps in AM reveal a key limitation: the combined effects of hysteresis, charge trapping, and ionic drift lead to progressive broadening and eventual loss of the Dirac peak. This degradation is particularly pronounced under electrolyte gating conditions. As shown in Fig. S6a, the Dirac peak is clearly visible during the initial sweeps but diminishes with continued operation. Although AM operation offers valuable insights into gate coupling and dielectric properties, its practical use in sensing is constrained by hysteresis-induced drift and the loss of peak resolution over time. To mitigate this issue and investigate AM further, we systematically performed dual-gate sweeps using aqueous electrolyte media (PBS, shown in Fig. 2b) by sweeping the top gate with fixed back gate potentials. We observed a shift in the Dirac peak with a slope of around $m = 13.6$ i.e. the ratio $C_{eff}^{TG}/C_{eff}^{BG}$, which is close to the theoretical prediction of 10 in Section 2. In order to explain this slight disparity from theory, we also measured the quantum capacitance of graphene using AM as described in Section S6 and found the value to be ~2 $\frac{\mu F}{cm^2}$ near the Dirac point increasing to $> 50 \frac{\mu F}{cm^2}$ farther from it [33]. This demonstrates that we expect to see a varied amplification factor >10 – as predicted in Section 2 – in cases where the quantum capacitance of graphene or the double layer capacitance is larger.

### Differential Feedback Modes: Differential Mode Sweep (DMS) and Differential Mode Fixed (DMF)

The Differential Feedback Mode represents a novel sensing architecture that leverages real-time electrostatic feedback to achieve intrinsic signal amplification. In Differential Mode Fixed (DMF) mode, the top gate is held at a fixed bias while the back gate is dynamically modulated via a closed-loop feedback mechanism implemented using off-the-shelf electronic components such as operational amplifiers or digital circuits (Fig. 2c). The op-amp senses the current flowing through the graphene channel and adjusts the back gate voltage ($V_{BG}$) to restore the channel to its predefined operating point. This enables the system to translate small changes in top gate potential – caused by molecular binding or environmental shifts – into amplified back gate voltage responses ($\Delta V_{BG}$), as described in Section 2. We compared different feedback modes as described in Section S7 and found that, while op amp-based feedback can effectively implement dynamic modulation, the high capacitance of the electrolyte can, in some cases, lead to instability in the op amp response. To mitigate this, in the present study we employed a "digital op amp" approach, implemented using a digital-to-analog converter (DAC) and analog-to-digital converters (ADCs), which eliminates the risk of feedback oscillations associated with the analog op amp configuration. We define a unifying parameter "Signal" for our tests as either $S = \Delta V_{BG}$ for differential mode measurements, $S = \Delta V_{\{BG,TG\},Dirac}$ for sweep methods, and $S = \Delta I_{ds} R_{gain}$ (gain of amplifier as discussed in Section S7) in case of static modes.

We expect an amplification factor of around 10-100× with PBS samples since the ratio $\frac{C_{eff}^{TG}}{C_{eff}^{BG}} \approx 10 - 100$ in our system (as discussed in Section 3.1). In this study, this factor is desirable since the molecules of interest in this study (e.g., redox-active neurotransmitters) typically generate $|\Delta V_{TG}| \leq 0.5V$ leading to $|\Delta V_{BG}| \leq$ ~15$V$ nearing the limits for safe wearable/portable battery sensors. Translating the signal into larger electrical shifts via feedback control not only improves signal-to-noise ratios but also simplifies the downstream analog-to-digital conversion and processing.

It is important to recognize that the amplification factor is tunable and dependent on the dielectric and electrochemical configuration of both gates. Should different classes of analytes or sensing environments be employed – e.g., those requiring organic solvents or non-polar media – the selection of the electrolyte (including top gate electrode) and back gate oxide must be carefully optimized. This tunability provides flexibility for tailoring the sensor design to the biochemical context, making this approach highly versatile.

In addition to signal amplification, we also investigated the effect of DMF on the intrinsic electrical noise characteristics of the GFET. Specifically, we measured the low-frequency 1/f noise – commonly associated with charge carrier fluctuations and interfacial traps – which is a major limiting factor in the detection of small signals. Our measurements revealed that DMF operation significantly suppresses the current noise compared to open-loop configurations such as TGF (As shown in Figure. S7). This noise suppression arises from the active stabilization of the channel current via the feedback loop, which dynamically compensates for fluctuations by modulating $V_{BG}$. As a result, the residual noise is effectively transferred from the current domain to the voltage domain, where it appears as minor variations in the amplified $V_{BG}$ signal.

In Differential Mode Sweep (DMS) mode, the top gate is swept while the differential feedback loop actively adjusts the back gate, enabling dynamic characterization of the graphene channel response under a range of electrochemical conditions as shown in Fig. 2d. This mode is particularly useful in assessing transient effects such as hysteresis, ionic relaxation timescales, and trap state filling dynamics, which are otherwise obscured in static or open-loop modes. Figure 2e plots the measured $V_{BG}$ against applied $V_{TG}$ in PBS and we find the slope (amplification factor) to be nearly ~10× confirming $\frac{C_{eff}^{TG}}{C_{eff}^{BG}} \approx 10$ in our system. Together, DMF and DMS constitute a powerful dual-gate framework with built-in electrostatic amplification, uniquely suited for low-signal environments.

## 4 Applications and Sensing Improvement across Various Analyte Classes

To benchmark the performance of various operational modes discussed in Section. 3 and highlight the superiority of DMF, we evaluated a range of biological, environmental, and chemical targets. This assessment quantifies improvements in key sensing metrics, including sensitivity, limit of detection (LOD), SNR, hysteresis, and drift.

### 4.1 pH

The detection of pH changes is a canonical application of liquid-gated GFETs due to the sensitivity of the electric double layer and the protonation/deprotonation of surface functional groups. In the single top gate (TGF/TGS) mode, pH sensitivity is manifested as a shift in the Dirac point voltage with increasing $H^+$ or $OH^-$ ion concentration. As shown in Fig. 3a, the pH sensitivity of TGF was 214 mV/pH and that of TGS exhibited significant drift (50.52%/hr; discussed in Section 4.6). Similarly, we calculated the sensitivity across different measurement modes described previously tabulated in Table. 2. We observed that dual-gate operation, particularly in differential feedback modes (DMF and DMS), enhances both the sensitivity and stability of pH sensing as shown in Fig. 3a. Dual-gate operation in DMF mode achieved a remarkable pH sensitivity of 1314 mV/pH, representing over a 6× amplification compared to single-gate modes – with a SNR of 21.64 and reduced drift of 0.66%/hr. These improvements are highlighted in Fig. 3a. The results, plotted in Fig. 3a and tabulated in Table 2, show that DMF operation reduced drift and hysteresis by more than an order of magnitude compared to TGS, confirming the theoretical predictions described in Section 2.

### 4.2 Redox-active Small Molecules

Many small molecules such as dopamine (DA), serotonin (SER), and norepinephrine (NEP), epinephrine (EP), uric acid (UA), and ascorbic acid (AA) are redox active, undergoing reduction–oxidation reactions at specific applied potentials [34–36]. These reactions modulate the interfacial charge density and, in turn, the electrostatic potential sensed by graphene. Consistent with our pH results, we observed a 21× amplification in sensor sensitivity for detecting these molecules, as shown in Fig. 3b. Owing to differences in the formal potentials of various redox-active neurotransmitters, the top-gate response is distinct for each analyte and is further amplified and measured using the DMF mode.

## 4.3 Protein Detection using Immobilized Capture Molecules

The detection of large biomolecules such as interleukin-6 (IL-6) is critically important in clinical diagnostics, particularly for inflammatory disorders, autoimmune conditions, and cancer progression monitoring. IL-6 is a cytokine with a molecular weight of ~21 kDa and a hydrodynamic diameter of approximately 4-6 nm. Unlike small molecules or ions, large biomolecules pose unique challenges GFETs due to limited charge transfer, slower diffusion kinetics, and steric hindrance near the graphene sensing surface.

To enable IL-6 detection, we functionalized the graphene surface with IL-6 antibodies, as described in Section 5.5 and subsequently tested the devices with increasing concentrations of IL-6. The results for TGF and DMF modes are shown in Fig. 3c. We observed that sensitivity is enhanced by 8× in DMF compared to TGF. Upon IL-6 binding, a net change in the local charge distribution and interfacial dipole potential shifts the Dirac point voltage, which in DMF mode was further amplified into a back-gate voltage response. These results, while demonstrated here for IL-6 detection, highlight a generalizable sensing paradigm applicable to a wide range of analytes and capture chemistries. The dual-gating strategy can, in principle, be extended to other clinically relevant targets such as cancer biomarkers (e.g., prostate-specific antigen, carcinoembryonic antigen), whole cells (e.g., circulating tumor cells, bacteria), and nucleic acids (DNA, RNA, microRNA). Likewise, the graphene transducer can be functionalized with diverse capture elements – including antibodies, aptamers, nucleic acids, and even intact cells – enabling both molecular and cellular detection in a label-free manner. Similar GFET-based devices have been reported for CRISPR-mediated nucleic acid detection with attomolar sensitivity via Cas-mediated cleavage products [37, 38], aptamer-functionalized GFETs for thrombin and cytokine sensing [39, 40], and antibody-modified GFETs for viral antigens and small-molecule toxins [41]. The generality arises from graphene's high interfacial sensitivity, the tunable surface chemistry for immobilizing different biorecognition layers, and the capability of the DMF mode to amplify surface-potential changes into measurable electronic signals.

## 4.4 Perfluoroalkyl and Polyfluoroalkyl Substances (PFAS)

Perfluoroalkyl and polyfluoroalkyl substances (PFAS) represent a critical class of persistent organic pollutants widely used in industrial and consumer products due to their hydrophobic and lipophobic properties. Their environmental persistence and bioaccumulation potential pose serious health risks, necessitating sensitive and field-deployable detection technologies. In this work, we demonstrated the applicability of dual-gated GFETs for detecting representative PFAS compounds

– specifically perfluorooctanoic acid (PFOA) as described in Section 5.5. The results are shown in Fig. 3d and tabulated in Table 2. TGF showed a sensitivity of 229 mV/dec with a LOD of 10 ppb and a SNR of 3.34, while DMF amplified the sensitivity to 3318 mV/dec, lowered the LOD to 1 ppb, and improved the SNR to 22.49. These results confirm the suitability of dual-gated GFETs for sensitive, real-time detection of PFAS contaminants in water.

### 4.5 Volatile Organic Compounds (VOC)

VOCs such as ethanol, acetone, and isopropyl alcohol are crucial indicators in environmental monitoring, breath analysis, and industrial safety. VOC molecules adsorb onto the graphene surface via physisorption or π-π interactions, modulating the carrier density and inducing a measurable shift in the Dirac point. The extent and direction of the shift depend on the electron-donating or withdrawing nature of the VOC; for example, 2-propanol typically acts as an electron acceptor, shifting the Dirac point toward more negative gate voltages.

We performed VOC sensing (gas environment) using BGF, BGS, and DMF modes, as described in Section 5.5 and Section S8. VOC detection was performed using BGF and DMF modes (Fig. 4a) and the performance metrics are tabulated in Table 2. BGF exhibited minimal drift (0.17%/hr; discussed in Section 4.6), but switching to BGS resulted in significant drift (38.91%/hr). Importantly, DMF maintained low drift (2.65%/hr) while still enabling active sensing. This demonstrates the advantage of DMF for stable, low-noise VOC detection.

### 4.6 Characterization of Signal Drift across Operation Modes

Long-term signal stability is a critical parameter in the design and operation of graphene-based chemical and biological sensors. Signal drift – which manifests as a gradual shift in the Dirac point or sensor output over time in the absence of analyte changes – can significantly impair sensitivity, reproducibility, and the reliability of quantitative measurements. Drift is often attributed to charge trapping/detrapping at the dielectric interface, ionic migration in the electrolyte, or hysteresis effects related to surface interactions. To evaluate and compare drift performance across different gating configurations, we systematically measured the time-dependent response of GFET sensors under both air-phase and liquid-phase conditions. The results are summarized in Fig. 4b–d.

Figure 4b presents the signal drift observed during VOC sensing using three different back-gate configurations BGF, BGS, and DMF. In the BGS mode, where the back-gate voltage is continuously swept during measurement, we observed significant signal drift – up to 38.91% per hour – which severely compromises signal stability and renders real-time sensing unreliable. In contrast, maintaining the back gate at a constant voltage in BGF mode drastically reduced drift to just 0.17% per hour, indicating that the sweeping process itself induces substantial disturbance, likely due to dynamic charge trapping in the gate dielectric or substrate, however in this mode SNR is lower. The DMF configuration, which incorporates active feedback from the source-drain current to dynamically adjust the back gate, showed a moderate drift of 2.65% per hour. While this is higher than the static BGF case, it represents a >15× reduction in drift compared to BGS, and crucially, it preserves the amplification gain described earlier in Section 2.

We also compared the drift if instead of $\Delta I_{DS}$ as the signal, we use $\Delta V_{Dirac}$ for the sweep modes, considering that the shift in Dirac peak location as the signal is very commonly sensor characterization [41]. The results for both top and back gate sweep modes are summarized in Fig. 4c, showing that the drift in $\Delta I_{DS}$ and $\Delta V_{Dirac}$ are comparable. For both $\Delta I_{DS}$ and $\Delta V_{Dirac}$ as the signal, the absolute drift is ~50% and ~40% in top gate and back gate sweep modes, respectively. One discernable difference observed is that while $\Delta V_{Dirac}$ drift is linear, the $\Delta I_{DS}$ drift is more non-linear. Nevertheless, this drift complicates sensor calibration and furthermore is dependent on the molecule/media under test, as well as any changes in the sweep window utilized [42].

Signal drift in electrolyte-gated modes – relevant for biological and chemical sensing in liquid environments – is shown in Fig. 4d for four configurations: TGS, TGF, DMF, and DMS. The TGS mode exhibited the highest drift at 50.52% per hour, consistent with the known instability of continuously swept electrolyte gates, where ionic redistribution and interfacial charge trapping dominate. TGF mode, where the electrolyte gate is held at a fixed potential, achieved the lowest drift, as expected, by minimizing ionic motion and electrostatic perturbation. Notably, the DMF mode reduced drift to just 0.66% per hour, representing a ~76× improvement over TGS, while still enabling dynamic control of the back gate and enhanced signal amplification. The DMS mode, where both gates are swept, produced a drift of ~7.03% per hour, which, while higher than DMF, still reflects a 7× reduction in drift compared to TGS.

The stark differences in drift performance across various operational modes highlight the critical role of gate modulation strategy in dual-gated GFET systems. In both air and liquid environments, gate voltage sweeps exacerbate charge trapping and hysteresis, particularly at the dielectric and graphene – electrolyte interfaces, ultimately degrading long-term reliability. By contrast, the active back-gate control in DMF mode provides real-time compensation of electrostatic fluctuations, dramatically stabilizing the sensor response. This enhanced drift suppression is key for enabling continuous, long-term monitoring in real-world sensing applications.

## 5 Materials and Methods

### 5.1 Materials

All chemicals and reagents were used as received without further purification. Phosphate-buffered saline (1× PBS, pH 7.4) was obtained from Thermo Fisher Scientific (Dulbecco's formulation without magnesium and calcium). Acetonitrile, BMIM (1-Butyl-3-methylimidazolium hexafluorophosphate) ionic liquid, serotonin (5-HT), epinephrine, norepinephrine, dopamine, ascorbic acid, uric acid, and perfluorooctanoic acid (PFOA), were purchased from Sigma-Aldrich. Interleukin-6 (IL-6) antibodies (monoclonal), bovine serum albumin (BSA), 1-ethyl-3-(3-dimethylaminopropyl) carbodiimide (EDC), and N-hydroxysuccinimide (NHS) were sourced from Thermo Fisher Scientific and Sigma-Aldrich. Monolayer graphene grown on copper foil was procured from Graphenea Inc. Silicon wafers (500 microns, P<100>, 0.001 – 0.005 ohm-cm) with thermally grown silicon dioxide (~285 nm) were purchased from Nova wafer Inc.

### 5.2 Fabrication of Substrates for Local Back Gating

Substrates for dual gating (with local back gate) were fabricated using standard photolithography and atomic layer deposition (ALD) techniques. A silicon wafer with a 285 nm thermally grown silicon dioxide layer was sequentially cleaned in acetone and isopropanol (IPA) and dried under nitrogen. A bilayer resist stack (LOR5A and SPR3012) was spin-coated and patterned using a Heidelberg MLA150 maskless aligner, and the exposed regions were developed in CD-26 developer to define the local gate areas. A tri-metal gate stack comprising 5 nm titanium (Ti), 20 nm gold (Au), and 10 nm platinum (Pt) was deposited via electron-beam evaporation. Lift-off was performed in acetone and PRS3000 remover, followed by oxygen plasma cleaning (25 W, 30 min; Harrick plasma cleaner) to remove residual organics and enhance surface hydrophilicity. Hafnium dioxide ($HfO_2$) was then deposited to a thickness of 35 nm by ALD (Lesker ALD150LE), and the thickness was confirmed using spectroscopic ellipsometry (J.A. Woollam). To expose the source and drain contact regions, a second photolithography step using the same bilayer resist stack was carried out, and the patterned $HfO_2$ areas were etched using reactive ion etching (PlasmaTherm Versalock) with an $Ar/CF_4$ gas mixture. The remaining resist was stripped in acetone, completing the fabrication of the local back-gate structure with $HfO_2$ dielectric.

### 5.3 Graphene Transfer and Device Fabrication

Graphene transfer was carried out using a poly(methyl methacrylate) (PMMA)-assisted wet transfer method. Commercial monolayer graphene on copper foil was spin-coated with PMMA and baked at 150°C for 2 minutes. To remove backside graphene, the reverse side was exposed to 25 W oxygen plasma in a Harrick cleaner for 15 minutes. Copper was then etched in ammonium persulfate solution (Transene company) and the PMMA/graphene stack was transferred onto the prepared gate stack (described in Section 5.2) through multiple rinsing in deionized (DI) water baths, followed by nitrogen drying.

To enhance adhesion of graphene to the gate-stack, the fabricated substrate was exposed to a brief 1-minute plasma treatment before picking up the floating film. Water was allowed to evaporate at 50°C for 15 minutes, followed by baking at 150°C for 15 minutes. The PMMA layer was removed by soaking in acetone for 4 hours, followed by a final bake at 200°C for 15 minutes to improve adhesion. To define the active graphene channel, SPR3012 resist was spin-coated, and photolithography was used to pattern non-channel regions. Exposed graphene was removed via plasma etching, and the resist was stripped by soaking in acetone for 4 hours. Finally, an SU-8 encapsulation layer was spin-coated, exposed, and developed to passivate the device and define the electrolyte-exposed sensing window. A commercial platinum wire electrode (BASi, Inc.) was used as the top gate.

### 5.4 Circuit Integration and Electrical Measurements

Each fabricated GFET was wire-bonded onto a custom-designed printed circuit board (PCB) that housed the measurement and control electronics. The system was managed by an Atmega2560 microcontroller. Analog signal conditioning was performed using TL074 operational amplifiers, with analog-to-digital conversion (ADC) handled by an MCP3204 module and digital-to-analog conversion (DAC) provided by an MCP4822 module. Relay switching between operational modes and grounding states was implemented using ULN2003A Darlington arrays, driven by GPIO lines expanded via an MCP23S08 I/O expander.

Each device could be independently switched to active or grounded states through relay control to minimize electrostatic discharge and leakage during inactive phases. Different biasing configurations (e.g., single-gate static, dual-gate dynamic) were established by selectively routing voltage and current lines through the relays. Drain current was sampled at 1 kHz over a 500 ms window and averaged to yield a steady-state value, while low-frequency noise characterization was performed by collecting 2-second current traces for offline analysis.

A custom graphical user interface (GUI), developed using the Panel Python library, enabled real-time data acquisition, device switching, and visualization. Backend communication with the circuit was implemented using Python and C++ serial protocols.

### 5.5 Measurement Procedures and Setups

For pH sensing, potassium chloride electrolyte in DIW was adjusted using hydrochloric acid (HCl) or sodium hydroxide (NaOH), and pH was verified using a VWR Versatile bench-top pH meter calibrated with standard buffer solutions. Each pH value was tested across 6 devices.

Small-molecule sensing (e.g., dopamine, serotonin, epinephrine, ascorbic acid, uric acid) was performed by preparing stock solutions in PBS and diluting serially to desired concentrations. Measurements were repeated on 6 devices per analyte to ensure reproducibility. Similarly, PFOA was tested with dilutions in DIW.

Volatile organic compound (VOC) detection was carried out in a custom-built sealed exposure chamber. Isopropanol was introduced via clean dry air (CDA) gas flow after aspirating a fixed volume of the liquid-phase VOC. Each exposure cycle was followed by purging and baseline recovery. Tests were conducted across 8 devices.

For IL-6 biosensing, graphene channels were functionalized using a carbodiimide crosslinking strategy. Briefly, pyrenebutyric acid (PBA) was adsorbed onto the graphene surface, followed by activation with EDC and NHS in MES buffer. Monoclonal IL-6 antibodies were then immobilized, and non-specific binding was blocked using 1% BSA in PBS. Devices were incubated with varying concentrations of IL-6 and washed before measurement. Each concentration point was tested across 6 devices.

## 6 Conclusion

In this work, we introduce a dual-gated graphene field-effect transistor (GFET) architecture that integrates a high-κ local back gate with an electrolyte top gate, enabling real-time, feedback-stabilized signal amplification for chemical and biological sensing. This asymmetric dual-gate design addresses longstanding limitations of single-gate GFETs – such as signal drift, hysteresis, and limited sensitivity – by eliminating the need for dynamic gate sweeps and reducing charge trapping and dielectric relaxation. Through systematic evaluation of multiple biasing modes across a broad set of analytes (pH, neurotransmitters, volatile compounds, environmental toxins, and proteins), we demonstrated that the Dual Mode Fixed configuration achieves superior performance, delivering >20× higher sensitivity, >7× noise reduction, and <15× drift compared to conventional methods. Maintaining constant transconductance via back-gate feedback enables over tenfold signal amplification while preserving a low-noise baseline, supporting robust, label-free detection in aqueous and physiologically relevant environments. Finally, the use of scalable materials and straightforward feedback electronics makes this platform readily adaptable to other 2D materials and miniaturized sensing technologies. Overall, this work defines a new operating paradigm for graphene-based sensors and provides a blueprint for designing high-performance, drift-resistant FET biosensors capable of real-time, multiplexed analysis in complex environments.

## Acknowledgment

The authors acknowledge partial support from NSF I/UCRC Phase II: Center for Atomically Thin Multifunctional Coatings (ATOMIC; Award #2113864), the NSF Division of Materials Research (DMR; Award #2323296), and the NSF Division of Electrical, Communications and Cyber Systems (ECCS; Award #2236997). The authors also thank the Roell Early Career Professorship Endowment for its support to A.E. V.K. further acknowledges the Center for Biodevices (CfB) at the Pennsylvania State University for the Leighton Riess Graduate Fellowship in Engineering.

## References

1. Kammarchedu V, Asgharian H, Zhou K, et al (2024) Recent advances in graphene-based electroanalytical devices for healthcare applications. Nanoscale 16:12857–12882. https://doi.org/10.1039/D3NR06137J

2. Ghosh R, Aslam M, Kalita H (2022) Graphene derivatives for chemiresistive gas sensors: A review. Mater Today Commun 30:103182. https://doi.org/10.1016/J.MTCOMM.2022.103182

3. Zhu J, Huang X, Song W (2021) Physical and Chemical Sensors on the Basis of Laser-Induced Graphene: Mechanisms, Applications, and Perspectives. ACS Nano 15:18708–18741. https://doi.org/10.1021/acsnano.1c05806

4. Liu J;, Bao S;, Wang X, et al (2022) Applications of Graphene-Based Materials in Sensors: A Review. Micromachines 2022, Vol 13, Page 184 13:184. https://doi.org/10.3390/MI13020184

5. Perala RS, Chandrasekar N, Balaji R, et al (2024) A comprehensive review on graphene-based materials: From synthesis to contemporary sensor applications. Materials Science and Engineering: R: Reports 159:100805. https://doi.org/10.1016/J.MSER.2024.100805

6. Bolotsky A, Butler D, Dong C, et al (2019) Two-Dimensional Materials in Biosensing and Healthcare: From *In Vitro* Diagnostics to Optogenetics and Beyond. ACS Nano 13:9781–9810. https://doi.org/10.1021/acsnano.9b03632

7. Ono T, Okuda S, Ushiba S, et al (2024) Challenges for Field-Effect-Transistor-Based Graphene Biosensors. Materials 2024, Vol 17, Page 333 17:333. https://doi.org/10.3390/MA17020333

8. Moldovan O, Iñiguez B, Deen MJ, Marsal LF (2015) Graphene electronic sensors – review of recent developments and future challenges. IET Circuits, Devices & Systems 9:446–453. https://doi.org/10.1049/iet-cds.2015.0259

9. Fu W, Jiang L, van Geest EP, et al (2017) Sensing at the Surface of Graphene Field-Effect Transistors. Advanced Materials 29:1603610. https://doi.org/10.1002/adma.201603610

10. Zeng Z, Wei W, Li B, et al (2022) Low Drift Reference-less ISFET Comprising Two Graphene Films with Different Engineered Sensitivities. ACS Appl Electron Mater 4:416–423. https://doi.org/10.1021/acsaelm.1c01066

11. Miyakawa N, Shinagawa A, Kajiwara Y, et al (2021) Drift Suppression of Solution-Gated Graphene Field-Effect Transistors by Cation Doping for Sensing Platforms. Sensors 2021, Vol 21, Page 7455 21:7455. https://doi.org/10.3390/S21227455

12. Mouro J, Domingues T, Pereira T, et al (2025) Analytical modeling and experimental characterization of drift in electrolyte-gated graphene field-effect transistors. NPJ 2D Mater Appl 9:26. https://doi.org/10.1038/s41699-025-00547-3

13. Wang GY, Lian K, Chu TY (2021) Electrolyte-Gated Field Effect Transistors in Biological Sensing: A Survey of Electrolytes. IEEE Journal of the Electron Devices Society 9:939–950. https://doi.org/10.1109/JEDS.2021.3082420

14. Saraswat V, Jacobberger RM, Arnold MS (2021) Materials Science Challenges to Graphene Nanoribbon Electronics. ACS Nano 15:3674–3708. https://doi.org/10.1021/acsnano.0c07835

15. Yang J, Jia K, Su Y, et al (2014) Hysteresis analysis of graphene transistor under repeated test and gate voltage stress. Journal of Semiconductors 35:094003. https://doi.org/10.1088/1674-4926/35/9/094003

16. Wang H, Wu Y, Cong C, et al (2010) Hysteresis of Electronic Transport in Graphene Transistors. ACS Nano 4:7221–7228. https://doi.org/10.1021/nn101950n

17. Zhao W, Zhang W, Chen J, et al (2024) Sensitivity-Enhancing Strategies of Graphene Field-Effect Transistor Biosensors for Biomarker Detection. ACS Sens 9:2705–2727. https://doi.org/10.1021/acssensors.4c00322

18. Xu H, Zhang Z, Xu H, et al (2011) Top-Gated Graphene Field-Effect Transistors with High Normalized Transconductance and Designable Dirac Point Voltage. ACS Nano 5:5031–5037. https://doi.org/10.1021/nn201115p

19. Kang JW, Cho WJ (2019) Achieving enhanced pH sensitivity using capacitive coupling in extended gate FET sensors with various high-K sensing films. Solid State Electron 152:29–32. https://doi.org/10.1016/J.SSE.2018.11.008

20. Sanjay S, Hossain M, Rao A, Bhat N (2021) Super-Nernstian ion sensitive field-effect transistor exploiting charge screening in WSe2/MoS2 heterostructure. npj 2D Materials and Applications 2021 5:1 5:1–8. https://doi.org/10.1038/s41699-021-00273-6

21. Knopfmacher O, Tarasov A, Fu W, et al (2010) Nernst Limit in Dual-Gated Si-Nanowire FET Sensors. Nano Lett 10:2268–2274. https://doi.org/10.1021/nl100892y

22. Le ST, Cho S, Zaslavsky A, et al (2022) High-performance dual-gate graphene pH sensors. Appl Phys Lett 120:. https://doi.org/10.1063/5.0086049

23. Smith C, Qaisi R, Liu Z, et al (2013) Low-Voltage Back-Gated Atmospheric Pressure Chemical Vapor Deposition Based Graphene-Striped Channel Transistor with High-κ Dielectric Showing Room-Temperature Mobility > 11 000 cm$^2$/V·s. ACS Nano 7:5818–5823. https://doi.org/10.1021/nn400796b

24. Liao L, Bai J, Qu Y, et al (2010) High-$\kappa$ oxide nanoribbons as gate dielectrics for high mobility top-gated graphene transistors. Proceedings of the National Academy of Sciences 107:6711–6715. https://doi.org/10.1073/pnas.0914117107

25. Konar A, Fang T, Jena D (2010) Effect of high-$\kappa$ gate dielectrics on charge transport in graphene-based field effect transistors. Phys Rev B 82:115452. https://doi.org/10.1103/PhysRevB.82.115452

26. Osenbach JW (1996) Corrosion-induced degradation of microelectronic devices. Semicond Sci Technol 11:155–162. https://doi.org/10.1088/0268-1242/11/2/002

27. Di Trani N, Silvestri A, Wang Y, et al (2020) Silicon Nanofluidic Membrane for Electrostatic Control of Drugs and Analytes Elution. Pharmaceutics 12:679. https://doi.org/10.3390/pharmaceutics12070679

28. Chonko MA (1988) Effects of Deionized Water Rinses on Gate Oxide Leakage Currents. In: The Physics and Chemistry of SiO2 and the Si-SiO2 Interface. Springer US, Boston, MA, pp 453–457

29. Kammarchedu V, Butler D, Rashid AS, et al (2024) Understanding disorder in monolayer graphene devices with gate-defined superlattices. Nanotechnology 35:495701. https://doi.org/10.1088/1361-6528/AD7853

30. Xia J, Chen F, Li J, Tao N (2009) Measurement of the quantum capacitance of graphene. Nat Nanotechnol 4:505–509. https://doi.org/10.1038/nnano.2009.177

31. Altman DG, Bland JM (2005) Standard deviations and standard errors. BMJ 331:903. https://doi.org/10.1136/BMJ.331.7521.903

32. Xia J, Chen F, Li J, Tao N (2009) Measurement of the quantum capacitance of graphene. Nat Nanotechnol 4:505–509. https://doi.org/10.1038/nnano.2009.177

33. Kim CH, Frisbie CD (2014) Determination of quantum capacitance and band filling potential in graphene transistors with dual electrochemical and field-effect gates. Journal of Physical Chemistry C 118:21160–21169. https://doi.org/10.1021/jp505391u

34. Mittal R, Debs LH, Patel AP, et al (2017) Neurotransmitters: The Critical Modulators Regulating Gut–Brain Axis. J Cell Physiol 232:2359–2372. https://doi.org/10.1002/JCP.25518

35. Sinha K, Das Mukhopadhyay C (2020) Quantitative detection of neurotransmitter using aptamer: From diagnosis to therapeutics. J Biosci 45:1–12. https://doi.org/10.1007/S12038-020-0017-X/FIGURES/3

36. Ribeiro JA, Fernandes PMV, Pereira CM, Silva F (2016) Electrochemical sensors and biosensors for determination of catecholamine neurotransmitters: A review. Talanta 160:653–679. https://doi.org/10.1016/J.TALANTA.2016.06.066

37. Guermonprez P, Nioche P, Renaud L, et al (2024) CRISPR–Cas Systems Associated with Electrolyte-Gated Graphene-Based Transistors: How They Work and How to Combine Them. Biosensors (Basel) 14:541. https://doi.org/10.3390/bios14110541

38. CHEN L, LI J, CHEN J, et al (2025) Advances in CRISPR-based gene editing technology and its application in nucleic acid detection. BIOCELL 49:21–43. https://doi.org/10.32604/biocell.2024.056698

39. Khan NI, Song E (2021) Detection of an IL-6 Biomarker Using a GFET Platform Developed with a Facile Organic Solvent-Free Aptamer Immobilization Approach. Sensors 21:1335. https://doi.org/10.3390/s21041335

40. Yu H, Zhao Z, Xiao B, et al (2021) Aptamer-Based Solution-Gated Graphene Transistors for Highly Sensitive and Real-Time Detection of Thrombin Molecules. Anal Chem 93:13673–13679. https://doi.org/10.1021/acs.analchem.1c03129

41. Sun M, Zhang C, Lu S, et al (2024) Recent Advances in Graphene Field-Effect Transistor Toward Biological Detection. Adv Funct Mater 34:2405471. https://doi.org/10.1002/adfm.202405471

42. Mouro J, Domingues T, Pereira T, et al (2025) Analytical modeling and experimental characterization of drift in electrolyte-gated graphene field-effect transistors. npj 2D Materials and Applications 2025 9:1 9:1–11. https://doi.org/10.1038/s41699-025-00547-3

**Table 1. Operational modes of the dual-gated GFET and their configurations**

| Mode Name | $V_{TG}$ | $V_{BG}$ | Signal |
|---|---|---|---|
| *Top Gate Fixed (TGF)* | Fixed | Floating | $\Delta I_{DS}$ |
| *Top Gate Sweep (TGS)* | Swept | Floating | $\Delta I_{Ds}$, $\Delta V_{TG,Dirac}$ |
| *Back Gate Fixed (BGF)* | Floating | Fixed | $\Delta I_{DS}$ |
| *Back Gate Sweep (BGS)* | Floating | Swept | $\Delta I_{Ds}$, $\Delta V_{BG,Dirac}$ |
| *Addition Mode (AM)* | Swept | Swept | $\Delta I_{Ds}$ |
| *Differential Mode Fixed (DMF)* | Fixed | Active | $\Delta V_{BG}$ |
| *Differential Mode Sweep (DMS)* | Swept | Active | $\Delta V_{BG}$ |

**Table 2. Summary of sensing metrics for various operational modes**

| Mode | Analyte | Sensitivity | LOD | SNR | Drift (%/hr) |
|---|---|---|---|---|---|
| TGF | pH | 214 mV/pH | | 5.71 | -0.22 |
| | Neurotransmitters | 1× | | 2.96 | |
| | IL-6 | 484 mV/dec | 50ng/mL | 6.80 | |
| | PFOA | 229 mV/dec | 10 ppb | 3.34 | |
| TGS | pH | | | | 50.52 |
| BGF | 2-Propanol (VOC) | 0.25 mV/ppm | | 6.13 | 0.17 |
| BGS | 2-Propanol (VOC) | | | | 38.91 |
| DMF | pH | 1314 mV/pH | | 21.64 | 0.66 |
| | Neurotransmitters | 21× | | 14.50 | |
| | IL-6 | 2588 mV/dec | 5ng/mL | 30.06 | |
| | PFOA | 3318 mV/dec | 1 ppb | 22.49 | |
| | 2-Propanol (VOC) | 14 mV/ppm | | 32.81 | 2.65 |
| DMS | pH | | | | -7.03 |

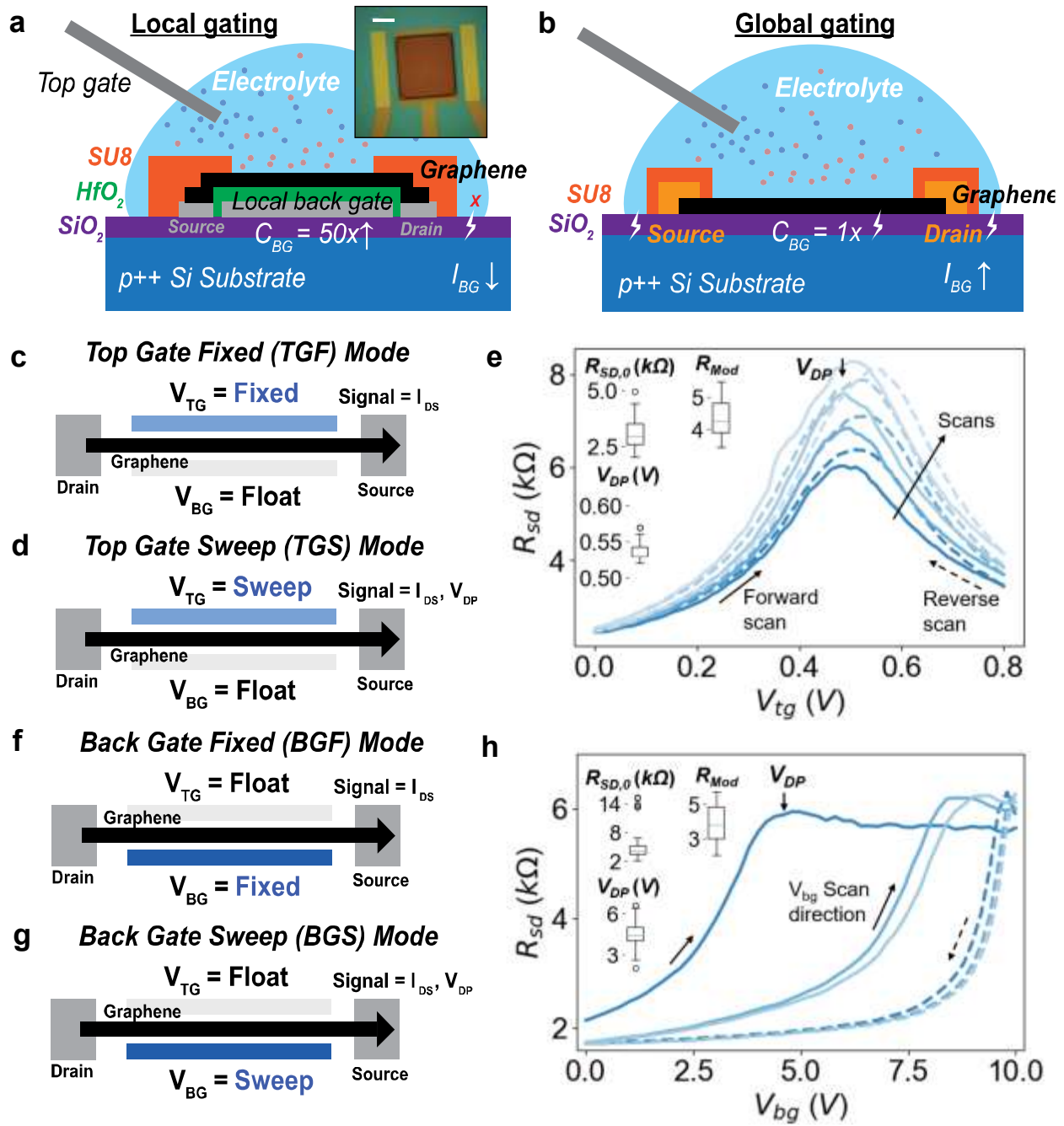

**Figure 1. Dual-gated graphene field-effect transistors (GFET) with one gate floated.** (a) Schematic of dual-gated GFET featuring an independent top gate and a locally patterned back gate with a solid-state HfO$_2$ dielectric, enabling capacitive modulation from both interfaces. Inset: optical micrograph of a fabricated device (scale bar: 30 μm). (b) Comparison schematic of a conventional global back-gated GFET with a thick SiO$_2$ dielectric. (c) Top Gate Fixed (TGF) mode: the top gate is biased while the back gate is floated. (d) Top Gate Sweep (TGS) mode: the top gate is swept with the back gate floated to obtain full transfer characteristics. (e) Representative transfer curves in TGS mode for aqueous PBS electrolyte media. (f) Back Gate Fixed (BGF) mode: the back gate is held at a fixed bias while the top gate is floated. (g) Back Gate Sweep (BGS) mode: the back gate is swept with the top gate floated to evaluate solid-state gating performance. (h) Representative transfer curves in BGS mode, demonstrating improved gating efficiency and lower voltage operation using the $HfO_2$ dielectric compared to conventional thick-$SiO_2$-based devices.

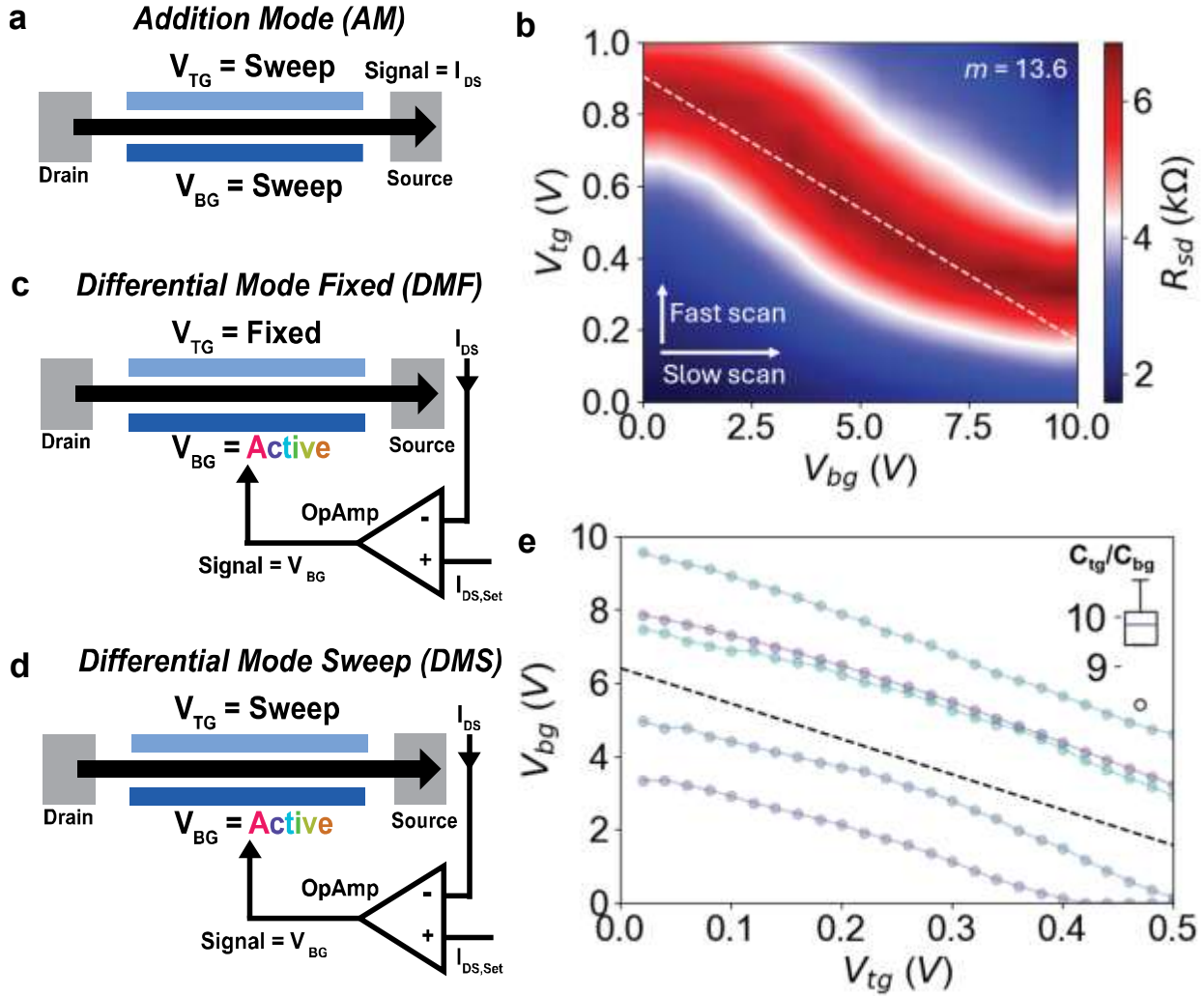

**Figure 2. Dual-gated operational modes in GFETs with both gates biased.** (a) Addition Mode (AM): sequential sweeping of both top liquid-gate and back solid-gate reveals distinct Dirac peaks for each gate. (b) 2D gate voltage maps obtained in PBS electrolyte showing the addition effect of dual gates. (c) Differential Mode Fixed (DMF): top gate is held constant while an operational amplifier adjusts the back gate in feedback to maintain a fixed channel current, enabling real-time signal amplification. (d) Differential Mode Sweep (DMS): top gate is swept while the back gate is dynamically adjusted, allowing investigation of transient and hysteresis effects. (e) Amplified back gate response DMS mode ($V_{BG}$) to the top gate sweep ($V_{TG}$) in PBS, with an observed slope ~10×, consistent with predicted feedback gain.

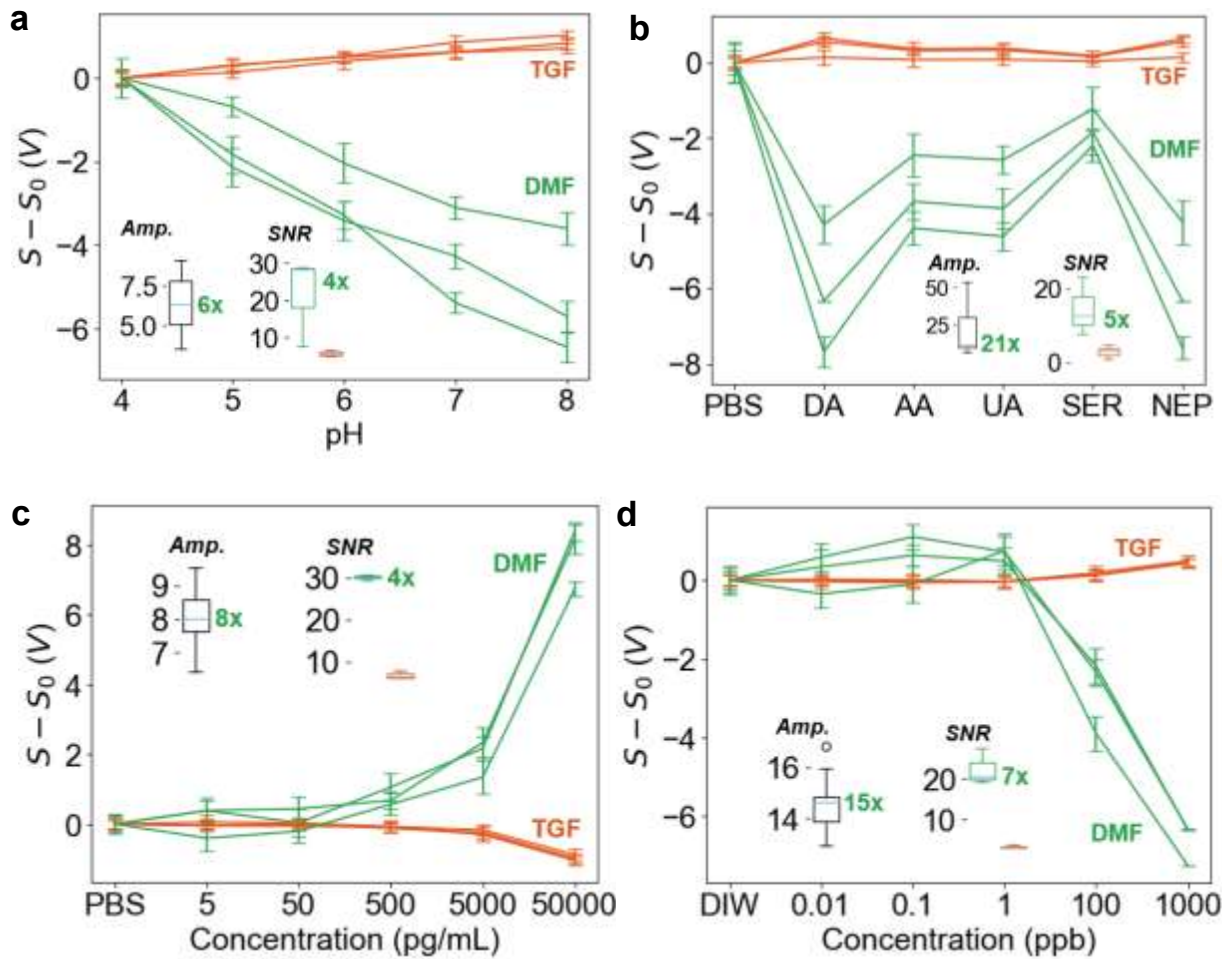

**Figure 3. Sensing performance of dual-gated GFETs for pH, redox-active small molecules, IL-6 (as a representative protein), and PFAS.** (a) pH sensing: comparison of top-gate mode (TGF) with differential feedback mode (DMF), showing enhanced stability and sensitivity in DMF. (b) Small molecule detection: sensing of redox-active small molecules demonstrates amplified response in DMF compared to TGF. (c) Protein detection (IL-6): IL-6 detection in TGF and DMF modes, with DMF providing higher sensitivity and stronger signal due to electrostatic feedback. (d) PFOA detection: comparison of TGF and DMF modes for PFOA sensing, highlighting improved response under DMF operation.

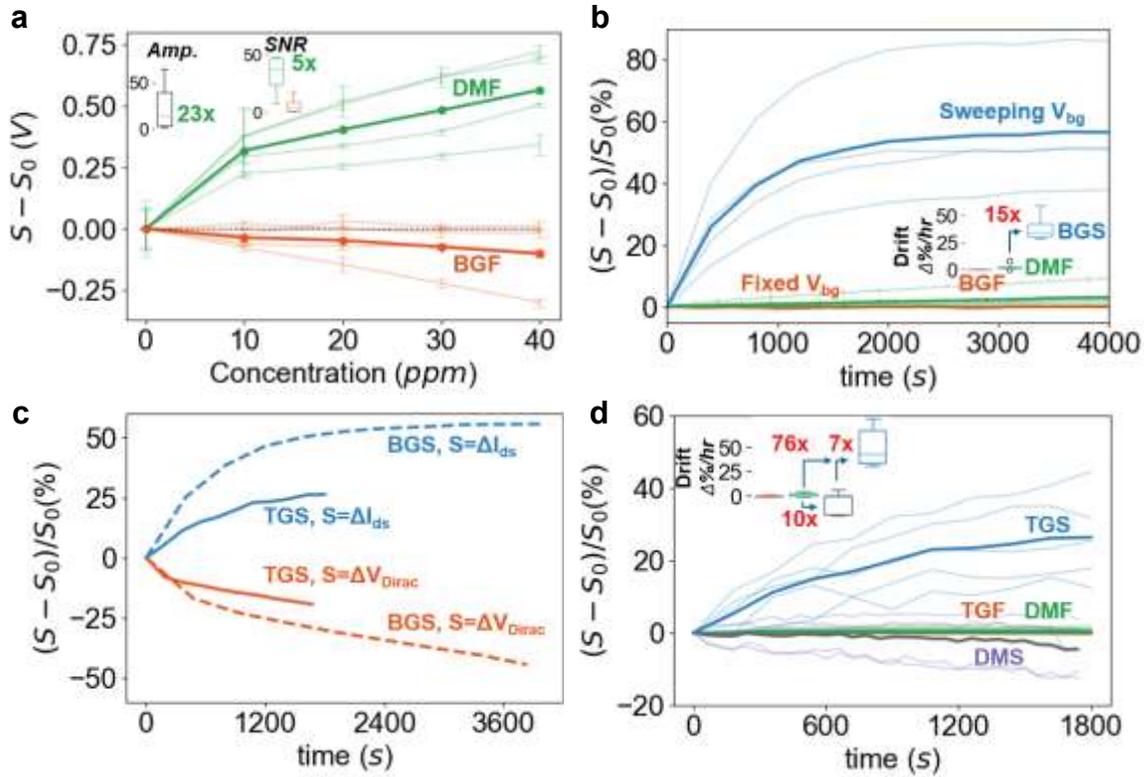

**Figure 4. VOC sensing and drift analysis.** (a) Detection of isopropyl alcohol using dual-gated GFETs operated in differential feedback mode (DMF) and back-gate fixed mode (BGF), showing enhanced signal response in DMF. (b) Signal drift comparison in BGF, DMF, and back-gate sweep (BGS). (c) Signal drift comparison in top and back-gate sweeps with current change and Dirac shift as the signals. (d) Signal drift comparison in top-gate modes: top-gate sweep (TGS), top-gate fixed (TGF), dual mode sweep (DMS), and DMF, highlighting the superior drift suppression achieved with DMF.

## S1. Fabrication of the local dual-gated graphene device

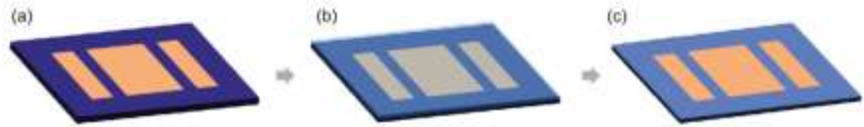

**Figure S1.** Fabrication process flow for the dual-gated substrate. (a) A tri-metal gate stack (5nm Ti / 20nm Au / 10nm Pt) is deposited via electron-beam evaporation to form the source drain contacts as well as the back gate metal. (b) 35 nm of $HfO_2$ is deposited by atomic layer deposition to form the gate dielectric. (c) Source/drain openings are etched using reactive ion etching to create the final structure showing the completed local back gate architecture.

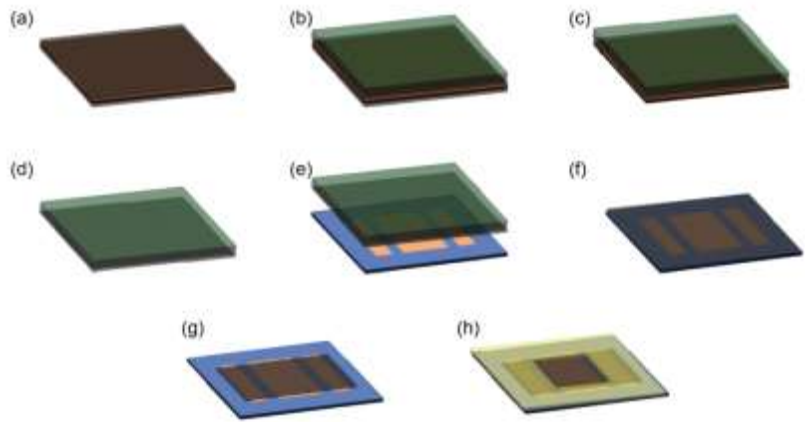

**Figure S2.** Key steps in the graphene transfer and device patterning process. (a) CVD Graphene on copper (b) PMMA-spin coating (c) Backside graphene etch (d) Copper etching (e) transfer to the target substrate (f) PMMA removal (g) Photolithographic patterning and plasma etching to define the graphene channel (h) SU-8 encapsulation to passivate the device and expose the sensing window for electrolyte gating

## S2. PCB and Software information

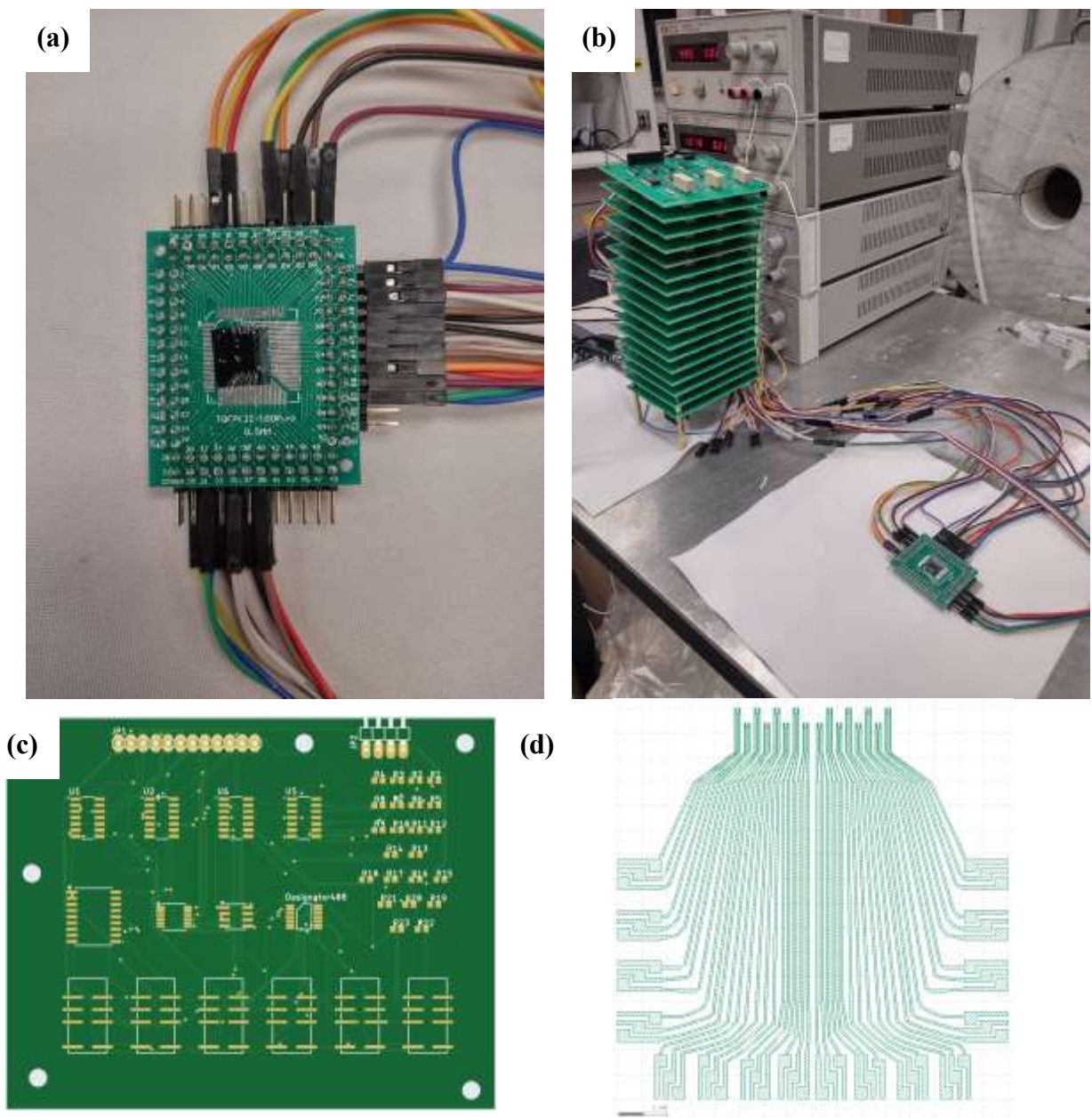

**Figure S3.** Multiplexed measurement setup for graphene sensor arrays. (a) Close-up image of a wire-bonded graphene device under test on the PCB platform. (b) Photograph of the stacked PCB system capable of simultaneously measuring up to 16 devices. (c) Schematic of the custom PCB design, illustrating the layout for individual device connections, signal routing, and integration with the measurement system. (d) Lithographic mask design of the 16-sensor array chip showing metal contacts to the sensors and contact pads.

## S3. Top Gate Sweep (TGS) data for different electrolyte media

Electrochemical impedance spectroscopy (EIS) was performed to characterize the double-layer capacitance and solution resistance of various electrolytes at the sensing interface. A three-electrode configuration was used, with a platinum counter electrode, Ag/AgCl reference electrode, and a gold working electrode. Measurements were carried out in the frequency range of 1 Hz to 1 MHz using a 50 mV AC perturbation at open circuit potential. The impedance spectra were fitted with a Randles equivalent circuit.

Theoretically, the back gate capacitance is calculated to be:

$$C_{bg} = \frac{\varepsilon_0 \varepsilon_r A}{d} \approx 38 \, pF$$

, where $\varepsilon_r = 25$ for hafnia, $d \cong 35 \, nm$, $A = 6000 \, \mu m^2$

And the corresponding geometric top gate capacitance from the $C_{dl}$ measurements with PBS are:

$$C_{tg} = C_{dl} \times A \approx 1388 \, pF$$

, where the top gate is exposed around $A \cong 3000 \, \mu m^2$

**Table S1. Media capacitance measurement using electrochemical impedance spectroscopy**

|  | $R_{sol}$ (k$\Omega$) | $C_{dl}$ @ 1Hz (F/cm$^2$) |
|---|---|---|
| DIW | 16.77 | 0.54p |
| 10mM KCl | 0.23 | 36.36u |
| PBS | 0.22 | 46.28u |
| ACET | 105.90 | 0.35p |
| ACET 10mM PHAS | 1.22 | 16.25u |
| IL | 4.83 | 8.51u |

## S4. Top Gate Sweep (TGS) data for all the devices tested

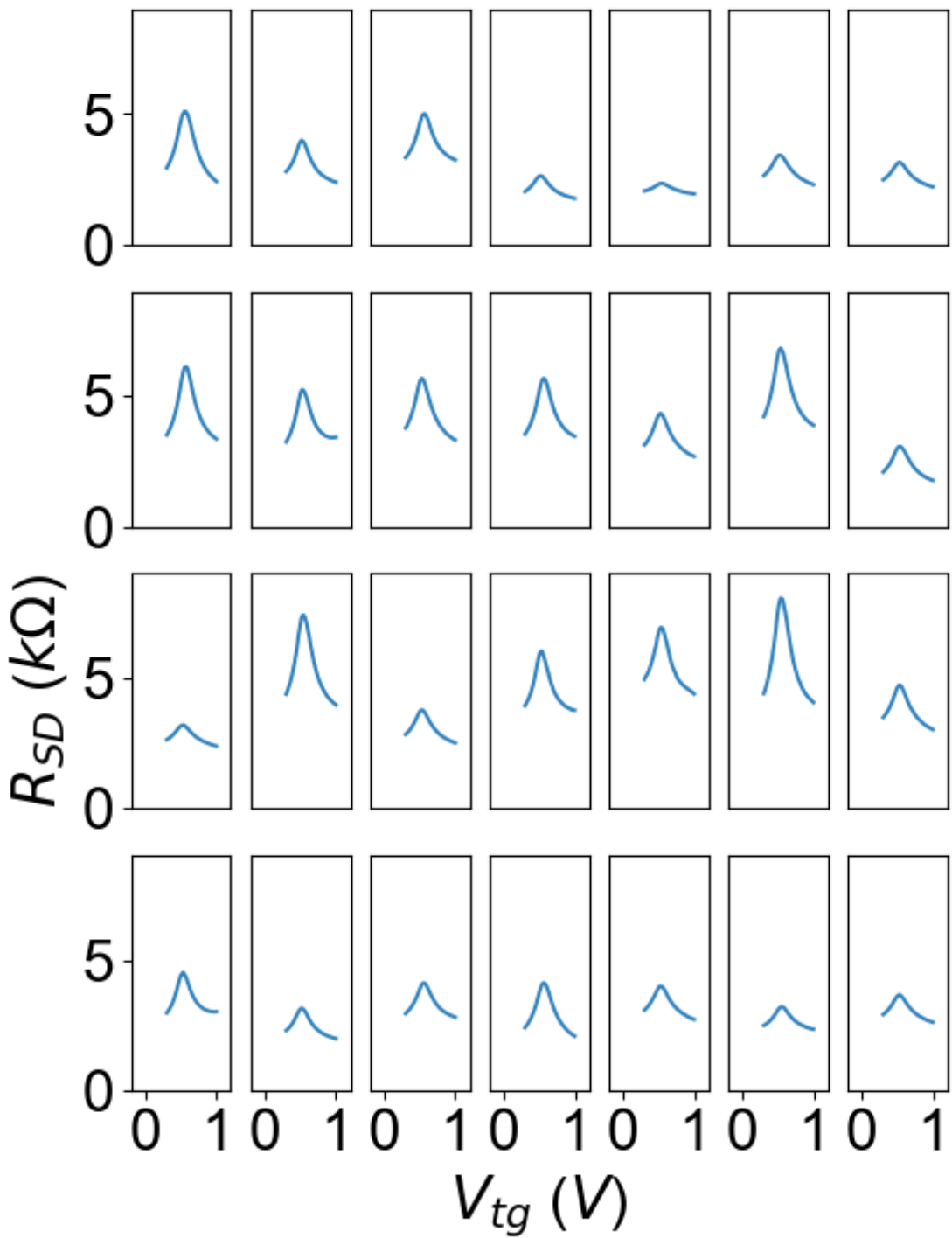

**Figure S4.** Top Gate Sweep (TGS) data for all the devices tested in PBS electrolyte. Raw data is available in the accompanying dataset for reference.

## S5. Back Gate Sweep (BGS) data for all the devices tested

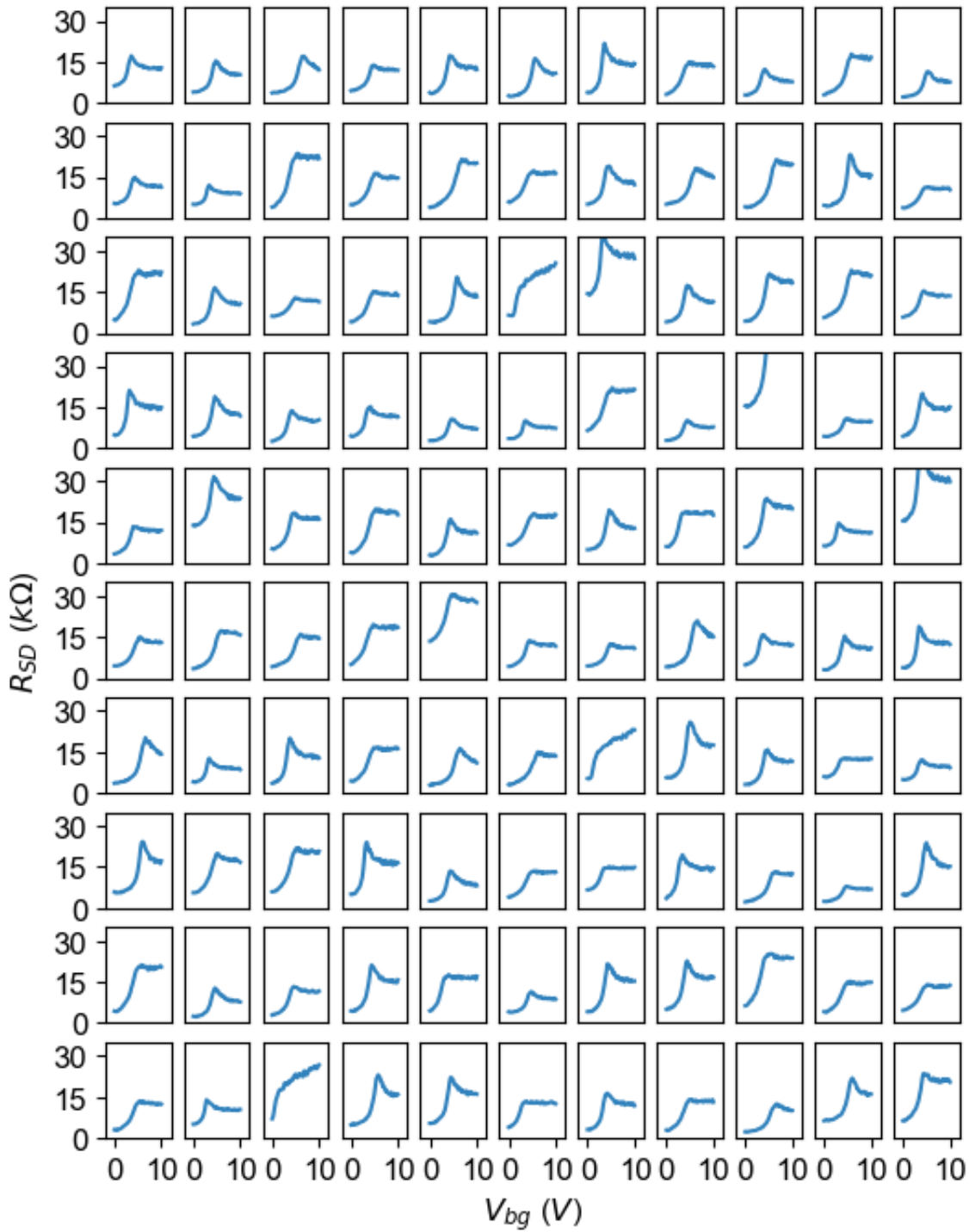

**Figure S5.** Back Gate Sweep (BGS) data for all the devices tested. Raw data is available in the accompanying dataset for reference.

## S6. Addition Mode (AM) data for different electrolyte media

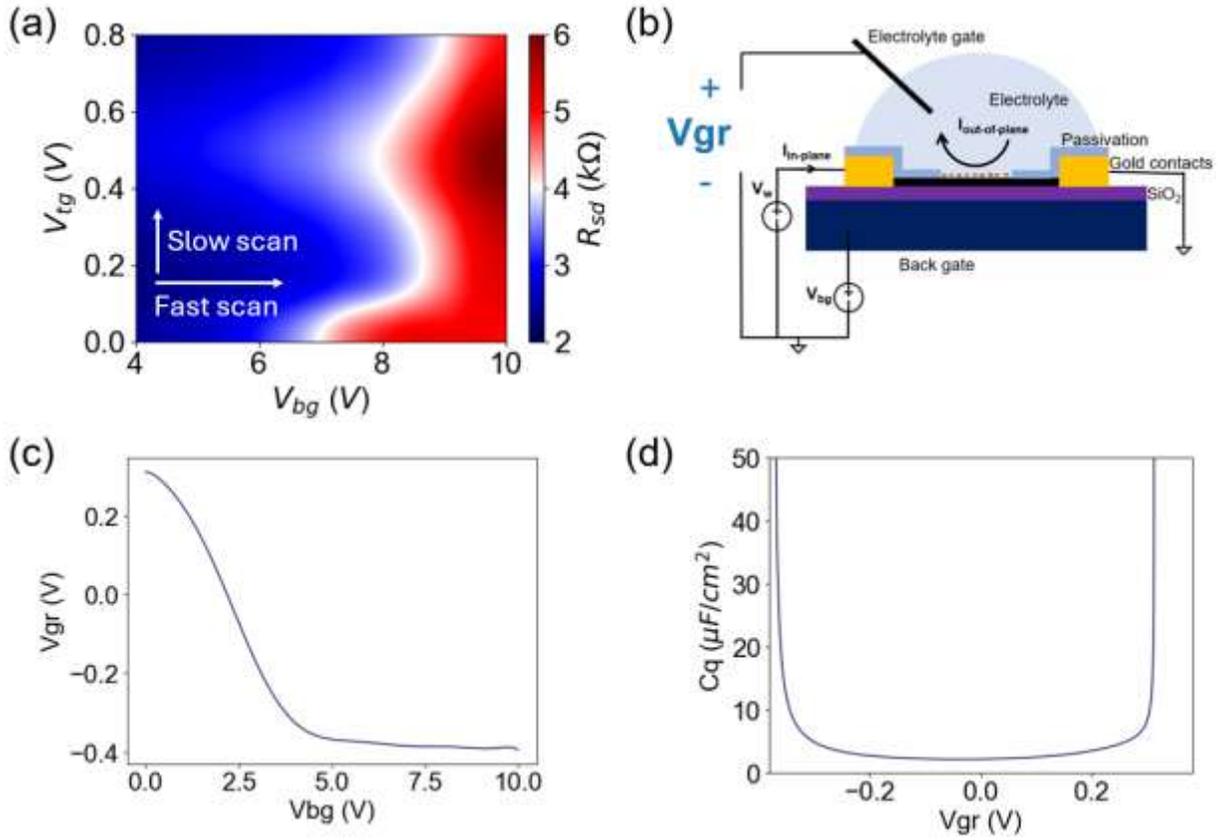

**Figure S6.** Dual-gated GFET: hysteresis and band-filling analysis. (a) Evolution of the Dirac peak in Addition Mode under repeated dual-gate sweeps. Early sweeps show a sharp charge-neutrality peak that broadens and shifts with subsequent measurements because of hysteresis, charge trapping, and ionic drift. (b) Schematic and definition of $V_{gr}$. During a back-gate sweep with a Pt top gate in electrolyte, $V_{gr}$ is the graphene potential measured versus a reference electrode in the electrolyte. (c) Measured $V_{gr}$ as a function of $V_{bg}$, reflecting the band-filling (Fermi-level) shift of graphene. (d) Quantum capacitance $C_q$ of graphene computed at different $V_{gr}$ values using the slope from panel (c) via $C_q = -C_{ox} \times (dV_{bg}/dV_{gr} + 1)$, showing a range of $C_q$ from about 2 microfarads per square centimeter to greater than 50 microfarads per square centimeter.

## S7. Differential Feedback Mode data

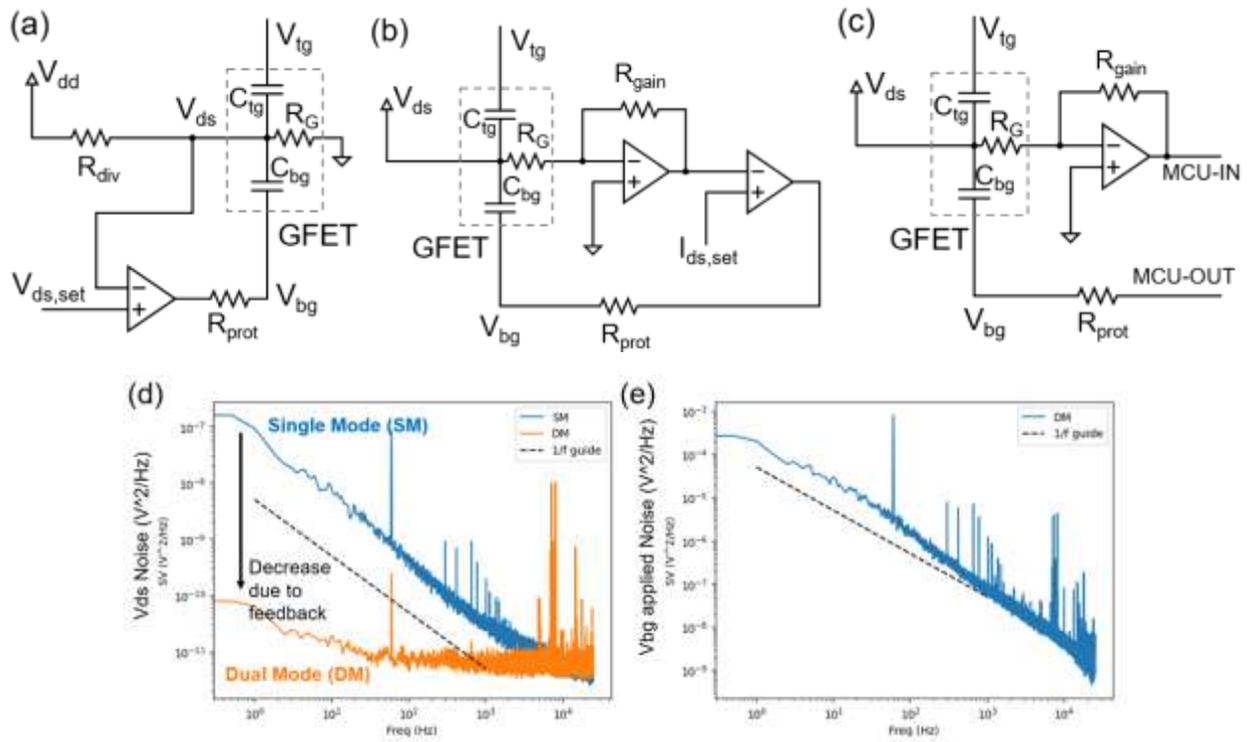

**Figure S7.** To evaluate the stability of different biasing strategies for dual-gated graphene field-effect transistors (GFETs), we simulated three representative circuit architectures (a-c) using an off-the-shelf operational amplifier model (TL074). The GFET was modeled as a gate-voltage-dependent channel resistance, denoted as $R_G$, varying from one kiloohm at $V_{bg}$ equal to zero volts to five kiloohms at $V_{bg}$ equal to five volts, with parasitic top-gate and back-gate capacitances, denoted as $C_{tg}$ and $C_{bg}$, respectively. $R_{gain}$ and $R_{prot}$ denote the trans-impedance gain and back-gate protection resistor respectively. (a) In this configuration, $V_{dd}$ is applied through a passive resistor network to the GFET channel, with $V_{ds}$ monitored at the divider midpoint. This approach is intrinsically stable due to its purely resistive bias path. However, due to high $V_{ds}$, gating is asymmetric across the channel introducing a differential gating effect, where both $V_{bg}$ and $V_{ds}$ influence the GFET's electrostatic state. This can shift the apparent transfer curve and complicate interpretation of sensing signals. (b) In this scheme, an operational amplifier maintains a constant $I_{ds}$ setpoint, denoted as $I_{ds,set}$, by adjusting $V_{bg}$ through feedback. While theoretically attractive for low $V_{ds}$ and automated bias control, the closed-loop dynamics involve multiple capacitances ($C_{tg}$ and $C_{bg}$) interacting with the op-amp's finite bandwidth. For realistic op-amp models such as TL074, the loop can approach instability in certain $R_G$-capacitance combinations. Compensation networks could improve stability but would add complexity to design and reduce bandwidth. Furthermore, with varying molecule/electrolyte, the response would be varied. For these reasons, this architecture was not adopted in the present work. (c) In this architecture, $V_{ds}$ is fixed at a low value (for example, ten millivolts), while $I_{ds}$ is measured via a shunt resistor and digitized at MCU-IN. Because $V_{ds}$ remains nearly constant regardless of $R_G$, the differential gating effect is greatly

suppressed, and the electrostatic control is dominated by $V_{bg}$. Due to digital control this method is stable over the full $R_G$ range, with minimal phase lag introduced by $C_{tg}$ and $C_{bg}$. It is therefore both effective in isolating back-gate control and straightforward to implement. (d) Reduction of current 1/f noise in the dual-mode feedback (DMF) system, specifically (a)-circuit. Spectral density of source-drain current noise measured under different gating modes, highlighting the suppression of low-frequency 1/f noise in the dual-mode configuration. (e) The corresponding noise transformed into equivalent back-gate voltage noise, demonstrating reduced gate voltage fluctuations in DMF. This reduction in both current and equivalent gate noise underscores the effectiveness of dual-mode feedback in stabilizing the device response and improving sensing resolution.

## S8. Volatile organic compound measurement setup

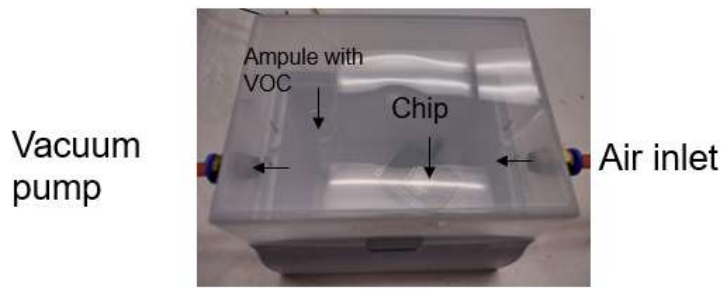

**Figure S8.** Optical image of the volatile organic compound (VOC) measurement setup. The system consists of an enclosed container purged with clean dry air (CDA) to establish a controlled baseline atmosphere. A known quantity of VOC is introduced by bubbling CDA through an enclosed container containing the liquid VOC, allowing controlled evaporation into the main chamber.